\documentstyle[prl,aps,twocolumn,epsf]{revtex}
\begin{document}
\twocolumn[\hsize\textwidth\columnwidth\hsize\csname@twocolumnfalse%
\endcsname
\title{Duality and the Modular Group in the Quantum Hall Effect}
\author{Brian P. Dolan}
\address{Department of Mathematical Physics, National University
of Ireland, Maynooth, Ireland}
\address{\rm and}
\address{Dublin Institute for Advanced Studies,
10 Burlington Rd., Dublin, Ireland}
\address{e-mail: bdolan@thphys.may.ie}
\date{Revised version: 25th December 1998}
\maketitle
\begin{abstract}
We explore the consequences of introducing a complex conductivity
into the quantum Hall effect. This leads naturally to an action of 
the modular group on the upper-half complex conductivity
plane. Assuming that the action of a certain subgroup,
compatible with the law of corresponding states,
commutes with the renormalisation group flow, we derive
many properties of both the integer and fractional
quantum Hall effects including: universality; the selection
rule $|p_1q_2-p_2q_1|=1$ for transitions between
quantum Hall states characterised by filling factors 
$\nu_1=p_1/q_1$ and $\nu_2=p_2/q_2$; critical values of
the conductivity tensor; and Farey sequences of transitions. 
Extra assumptions about the
form of the renormalisation group flow lead to the
semi-circle rule for transitions between Hall plateaux.

\end{abstract}
\bigskip
\hfill{PACS nos: 73.40.Hm, 05.30.Fk, 02.20.-a} 
\bigskip
]

The purpose of this letter is to explore the consequences of the proposal,
made in
\cite{LutkenRossB} and examined further in \cite{Lutken} \cite{BL},
that the hierarchical structure of the zero temperature integer and
fractional
quantum Hall effects can be interpreted in terms of the properties
of a subgroup of the modular
group, $Sl(2,{\bf Z}):=\Gamma(1)$ --- specifically 
the subgroup which consists of elements of
$\Gamma(1)$ whose bottom left entry is even, sometimes denotd
$\Gamma_0(2)$ in the
mathematical literature.  This group acts on the upper-half complex
plane, parameterised by the complex conductivity,
$\sigma=\sigma_{xy}+i\sigma_{xx}$, in units of ${{e^2}\over{h}}$,
and is generated by
two operations, $T:\sigma\rightarrow\sigma + 1$ and
$X:\sigma\rightarrow{{\sigma}\over{2\sigma + 1}}$.  If
$\gamma =\pmatrix {a&b \cr
2c&d}\in\Gamma_0(2)$, with $a,b,c,$ and $d \in {\bf Z}$ and $ad-2cb=1$,
then
$\gamma(\sigma)={{a\sigma +b}\over{2c\sigma + d}}$.  Thus
$T=\pmatrix{1&1\cr 0&1}$
and $X=\pmatrix{1&0\cr 2&1}$.
Some consequences of this assumption for the phase diagram in the 
$\sigma$-plane were examined in \cite{Lutken} and in the second of 
these references the author notes that there is a connection
with the work of Kivelson, Lee and Zhang \cite{KLZ}, but remarks that
the comparison between \cite{Lutken} and \cite{KLZ} is not
immediate. One of the aims of this paper is to explore the relation
between these two approaches.
\par
Following \cite{LutkenRossB}---\cite{BL}, 
it will be assumed that the phase diagram for the
quantum Hall effect
can be generated by the action of $\Gamma_0(2)$ on the upper-half
$\sigma$ plane. 
This immediately implies the \lq law of corresponding states'
of \cite{KLZ} 
and \cite{LKZ}. 
At Hall plateaux we have $\sigma_{xx} = 0$ and $\sigma_{xy} = s$ where $s$ is a
ratio of two mutually prime integers, with odd denominator (note that $s$ is being used
here to label the quantum phases and is denoted by $s_{xy}$ in \cite{KLZ}).
Plateaux can be related to each other
by repeated
action of $T$ and $X$. At the center of the plateaux,
the filling factor, $\nu$, is equal to the ratio $s=p/q$ and 
$T: \nu\rightarrow \nu + 1$ is the Landau
level addition
transformation of \cite{KLZ} while $X: \nu\rightarrow
{{\nu}\over{2\nu+1}}$ is the flux
attachment transformation.  The particle-hole transformation
$\nu\rightarrow 1 - \nu$,
can be realised as the outer auto-morphism $\sigma\rightarrow 1 - 
\bar \sigma$
acting on the upper-half plane, where $\bar\sigma=\sigma_{xy}-i\sigma_{xx}$
(it will be assumed throughout that the electron spins are split,
for the spin degenerate case one must re-scale $\sigma\rightarrow 2\sigma$).
\par
The upper-half $\sigma$-plane can be completely covered by copies of a
single \lq
tile\rq , or fundamental region (see e.g. \cite{Rankin}),
related to each other by elements of
$\Gamma_0(2)$.  The fundamental region has cusps at 0 and 1,
linked by a semi-circle of unit diameter, and consists of a vertical strip
of unit width constructed above this semi-circle.  
By assumption all allowed quantum Hall
transitions are images of the transition $\nu=0\rightarrow \nu=1$ under some 
$\gamma \in \Gamma_0(2)$, and
hence also linked by a semi-circle.  
\par
Each such semi-circle has a special point, in addition to the end points,
which is a fixed
point of $\Gamma_0(2)$
in the following sense.
The point $\sigma^\ast = {{1+i}\over{2}}$
is left fixed by $\gamma^\ast=\pmatrix{1&-1\cr 2&-1}$.
Similarly the points obtained from $\sigma^\ast$ by the other
elements of $\Gamma_0(2)$, $\sigma_\gamma^\ast:=\gamma(\sigma^\ast)$,
are left fixed by $\gamma\gamma^\ast\gamma^{-1}$.
It can be shown that the imaginary part, $\Im (\sigma_\gamma^\ast)
\leq {1 \over 2}$
or $\Im (\sigma_\gamma^\ast)=\infty,\; \forall\gamma$.
The points $\sigma_\gamma^\ast$ can be interpreted as critical points for the
transition $\gamma(0)\leftrightarrow\gamma(1)$ if we further assume that
the action of
$\Gamma_0(2)$ commutes with the renormalisation group (RG) flow.  For if
$\sigma_\gamma^\ast$ were not a RG fixed point, we could move to an
infinitesimally close point $\phi(\sigma_\gamma^\ast) \neq
\sigma_\gamma^\ast$
with a RG transformation, $\phi$.  Demanding $\gamma \circ
\phi(\sigma_\gamma^\ast)=\phi
\circ\gamma(\sigma_\gamma^\ast)=\phi(\sigma_\gamma^\ast)$
then implies that
$\phi(\sigma_\gamma^\ast)$ is also left invariant by $\gamma$. But the 
fixed points of $\Gamma_0(2)$ are isolated, so there
is no other fixed point infinitesimally close
to $\sigma_\gamma^\ast$.  Hence
$\phi(\sigma_\gamma^\ast)=\sigma_\gamma^\ast$ and
$\sigma_\gamma^\ast$ must be a RG fixed point.
The end points of the arches, at $\sigma=\nu$ with $\nu=p/q$ rational,
are also fixed points of $\Gamma_0(2)$. For $q$ odd these are
stable Hall states.
Note, however that a fixed point of the
RG need not necessarily be a fixed point of
$\Gamma_0(2)$ --- but there is no experimental
evidence of such extraneous fixed points of the RG.
\par
Thus the fixed points of $\Gamma_0(2)$ must be fixed points of the
RG, i.e. critical points.  This leads to the 
topology of the
flow diagram of
\cite{Lutken}, reproduced here in figures 1 and 2 where solid
lines represent phase
boundaries and dashed lines represent quantum Hall transitions.
This implies  the flow diagram proposed in \cite{Khem},
with its experimental support \cite{WCTPR} and is also compatible with the
phase diagram
derived in \cite{KLZ}.
That $\sigma^\ast = {{1+i}\over{2}}$ is a
critical point for the lowest Landau level was argued in \cite{HHB}. 
Phase boundaries
and transitions are represented by semi-circles in the
figures, but this is not forced
on us by the $\Gamma_0(2)$ symmetry. They could be
distorted from this geometry,
provided that all phase boundaries are copies of a distorted
\lq fundamental' phase boundary (running
from ${1\over 2}$ to ${1\over 2}+i\infty$) under the
action of $\Gamma_0(2)$. Similarly
the dashed transition trajectories must all be copies of a distortion of the
\lq fundamental' arch spanning $0$ and $1$. Note, however that the
{\it fixed
points are immovable}. A useful aspect of the semi-circular arches used in
the figures 
is that the intersection of any solid phase boundary with a dashed transition
is a fixed point of $\Gamma_0(2)$, as are the end points of the arches (which
are rational numbers or points at $\sigma = r+i\infty$ for integral or
half-integral $r$). 
Any distortion from semi-circular geometry must leave the end points and
intersections of phase
boundaries and transition trajectories pinned at the fixed points of
$\Gamma_0(2)$.
  
\par
As in \cite{KLZ}, the phase diagram generated by $\Gamma_0(2)$ determines
which
transitions are allowed and which are not.  Thus, for example, $s: {1\over
3}\rightarrow 0$ is allowed while $s: {1\over
3}\rightarrow {1\over 7}$ is not.  All allowed
transitions are generated by
acting on the arch passing through $\sigma=0$
and $\sigma=1$
by some, $\gamma\in\Gamma_0(2)$. This allows the derivation a
selection rule for
 a transition $s_1=p_1/q_1 \rightarrow
s_2=p_2/q_2$, where
$q_1$ and $q_2$ are odd, and the pairs $p_i$ and
$q_i$ ($i=1,2$) are
relatively prime (for brevity we shall not always distinguish below
between $s$, labelling the quantum Hall phase, and $\nu$, the
filling factor, except where necessary for comparison with \cite{KLZ}
--- on the real axis, when $\sigma_{xx}=0$, they are the same).  We shall see that a transition is allowed if and
only if  $p_1q_2 -p_2q_1 = \pm 1$.  

From the above assumptions we
have (relabeling if
necessary) $\nu_1 = \gamma(0)$, $\nu_2=\gamma(1)$.
Thus ${{p_1}\over{q_1}}={b\over d}$
and ${{p_2}\over{q_2}}={{a+b}\over{2c+d}}$ where
$\gamma=\pmatrix{a&b\cr 2c&d } \in
\Gamma_0(2)$.  
Since $ad-2cb=1$, $b$ and $d$ are mutually
prime, by an elementary
result of number theory, hence (taking plus signs without
loss of generality)
$b=p_1, d=q_1$. 
Similarly $(a+b)d-(2c+d)b=1$ implies that $a+b$ and $2c+d$
are mutually prime, hence $a+b=p_2$ and $2c+d=q_2$.
Thus $\gamma=\left(\matrix{p_2-p_1 & p_1 \cr q_2-q_1 & q_1}\right)$
and the condition $det\gamma=1$ then requires $p_2q_1-p_1q_2=1$. 
The only possible exception to
this rule would be a
transition from $\nu= n\rightarrow \nu=m$ ($n,m\in{\bf Z}$), which could
occur by going first
from $\sigma=n$ to $\sigma=n+i\infty$ and then in to $\sigma=m$
from $\sigma=m+i\infty$.
In a real experiment the maximum value of
$\vert \sigma \vert$
would presumably be finite, due to impurities.
\par
One can determine sequences of allowed transitions as follows.  Suppose
$\nu_0=p_0 /q_0$, with $q_0$ odd, is an allowed state, with $p_0$ and
$q_0$ relatively prime. Consider the sequence
$\nu_n={{kn+p_0}\over{ln+q_0}}:={{p_n}\over{q_n}}$ for $n,k,l
\in {\bf Z}$,
where $l$ is even (so that $q_n$ is odd).
Then $p_{n+1}q_n-p_nq_{n+1}=\pm 1 \Leftrightarrow kq_0-lp_0 =\pm 1$. Thus a
transition $\nu_{n+1}\rightarrow \nu_n$ is allowed provided $\vert kq_0-lp_0
\vert = 1$.  In this way we can, for example, generate the three sequences
$$
{1\over 3}\rightarrow{2\over 5}\rightarrow{3\over
7}\rightarrow{4\over
9}\rightarrow{5\over 11}\rightarrow{6\over 13}\rightarrow
\dots $$
$$\dots \rightarrow {7\over 13}\rightarrow{6\over
11}\rightarrow{5\over
9}\rightarrow{4\over 7}\rightarrow{3\over
5}\rightarrow{2\over 3}\rightarrow 1$$
$$
{2\over 3} \rightarrow {5\over 7}\rightarrow{8\over
11}\rightarrow{11\over
15}\rightarrow \dots
\eqno (1)$$
plus higher sequences obtained by adding an integer to each term in these
sequences. Such sequences are called Farey sequences and their relevance to
the quantum 
Hall effect was examined in \cite{ZB}.
Note that a given experiment may jump from one sequence to another.
Thus
$$
\dots \rightarrow {3\over 5}\rightarrow{2\over 3}\rightarrow{5\over
7}\rightarrow\dots
$$
is observed in \cite{WESTGE}.
\par
Each transition contains a critical point given by
$\gamma(\sigma^\ast)$.  Thus if
$\gamma = \pmatrix{a&b\cr 2c&d}$, the critical point is at
$$
\sigma^\ast_\gamma={{2ac+2bc+ad+2bd+i}\over{2d^2+4cd+4c^2}}=
{(p_1q_1+p_2q_2)+i\over (q_1^2+q_2^2)}\eqno(2)
$$
when the transition goes from $\nu_1=\gamma(0)=b/d=p_1/q_1$ to
$\nu_2=\gamma(1)={{a+b}\over{2c+d}}=p_2/q_2$. The parameters of $\gamma$
can be related to
physical parameters as follows. Following \cite{LKZ}, let $\eta$ be the
effective charge of
the quasi-holes of a Hall state, $e^\ast=\eta$, 
$\theta$ the statistical parameter (i.e. the phase of the two quasi-particle
wave 
function changes by $\pi\theta$ when the positions of the two particles are
exchanged)
and $s$ be the Hall state, with magic filling factor $\nu=s$.
Then the critical conductivity for a
transition from $s=\nu$ to $s^\prime=\nu-\eta^2/ \theta$ is given by equation (26) of
\cite{LKZ} (in
dimensionless units)
$$\sigma_{xx}={{\eta^2}\over{1+\theta^2}},\qquad
\sigma_{xy}=s-\theta{{\eta^2}\over{1+\theta^2}}\eqno (3)
$$
Equating these with the critical values in equation (2),
there are two
possibilities, depending on whether $\nu=\gamma(1)$ or
$\gamma(0)$, 
$$
\hbox {i)}\qquad \nu={a+b\over 2c+d}={p_2 \over q_2} \quad ,\,\,\,
\theta={d\over 2c+d}={q_1 \over q_2},
$$
$$\eta^2={1\over (2c+d)^2}={{1}\over{q_2^2}} \eqno (4) 
$$
$$
\hbox{ii)} \qquad \nu={b\over c}={p_1\over p_2} \quad , \,\,\,
\theta=-{(2c+d)\over d}=-{q_2\over q_1},
$$
$$\eta^2= {1\over d^2}={1\over q_1^2}. \eqno
(5)
$$
In both cases we reproduce the result, that $\eta=\pm 1/q$,
\cite{LH}
and \cite{CMH2F}.
Note in passing that the transition from bosonic to
fermionic conductivities given by 
equation (14) of reference \cite{KLZ} is implemented by the
action of an element of $\Gamma(1)$
which is not in $\Gamma_0(2)$. Thus $\sigma=\gamma(\sigma^{(b)})$
where $\sigma^{(b)}=\sigma^{(b)}_{xy}+i\sigma^{(b)}_{xx}$ is the
complex conductivity of the bosonic Chern-Simons
action
and $\gamma={1\over\eta}\pmatrix{\eta^2-\theta s &s\cr
     -\theta&1\cr}$.
The above discussion gives the explicit connection between the
Chern-Simons analysis of \cite{KLZ} and the group theory
analysis of \cite{Lutken}. 
\par
We make a final comment about the \lq semi-circle' law
of reference \cite{DR} - \cite{RCH}.
By assumption, each quantum Hall 
transition can be obtained from the one between 0 and 1, passing through
$\sigma^\ast={{1+i}\over{2}}$, by the action of some element of
$\Gamma_0(2)$. Since
$\Gamma_0(2)$ maps semi-circles built on the real axis into other
such semi-circles
we can deduce the \lq semi-circle law\rq $\,\,$ of reference
\cite{DR} - \cite{RCH} by making one extra assumption --- that the
\lq fundamental' arch
between $0$ and $1$ is a semi-circle. This implies that
{\it all} other
transitions
are semi-circles and allows predictions to be made of the
maximum values of $\sigma_{xx}$
and $\rho_{xx}$ in any allowed transition, $\nu_1\rightarrow \nu_2$,
as well as the
values of $\sigma_{xy}$ and $\rho_{xy}$ at which they occur.
Thus the maximum
value of $\sigma_{xx}$ is at
$\sigma_{xx}^{max}={{\nu_1-\nu_2}\over{2}}$, where
$\sigma_{xy}={{\nu_1+\nu_2}\over{2}}$ (where $\nu_1>v_2)$.
In general, this does not
coincide with the critical value
$\sigma_\gamma^\ast = \gamma(\sigma^\ast)$,
except for the integer transitions (table 1).
\par
The maximum value of $\rho_{xx}$ is found by constructing the
semi-circle
through ${1\over \nu_1}$ and ${1\over \nu_2}$ (provided neither
vanishes).  Thus
$\rho_{xx}^{max}={1\over 2}\bigl({1 \over \nu_2} -
{1\over \nu_1}\bigr)$, where
$\rho_{xy}={1\over 2}\bigl( {1\over \nu_2} + {1\over \nu_1}\bigr)$.
Some
representative examples are shown in table 1.
\par
To summarise, assuming (as in \cite{BL}) that the phase and
flow diagram for the
upper-half complex conductivity plane can be generated from an
action of
$\Gamma_0(2)$ which  commutes with the RG, one deduces:
(i) that all critical points are given by
$\gamma\bigl(\sigma^\ast\bigr)$, where $\sigma^\ast={{1+i}\over{2}}$, 
with $\gamma\in\Gamma_0(2)$; (ii)
the phase diagram of \cite{KLZ}, \cite{Lutken}  and \cite{WCTPR};
(iii) the laws of corresponding states \cite{KLZ}, \cite{LKZ};
and (iv) the selection rule $|p_1q_2-p_2q_1|= 1$, 
dictating which transitions
are allowed and which are forbidden.  Lastly, the semi-circle
law of \cite{DR} - \cite{RCH} is
compatible with, but not implied by, $\Gamma_0(2)$.
\par
It should be noted that the full modular group does {\it not}
provide the
correct phase structure, since it would imply further critical
points at the images
of $\sigma=i$ and $\sigma={{1+i\sqrt 3}\over{2}}$, under
$\gamma\in \Gamma(1)$, which
are not observed experimentally. The appearance of
$\Gamma_0(2)$ is due to the
extension of Kramers-Wannier duality $\sigma_{xx}\rightarrow 1/
\sigma_{xx}$ to the
whole complex plane.  It was argued in \cite{CR} that this
extension leads naturally to
$\Gamma(1)$ acting on the upper-half complex plane, for a
coupled clock model. This
was applied to the quantum Hall effect in 
\cite{SW} and \cite{LR}.  It appears to have been
noted first in \cite{LutkenRossB} that the subgroup $\Gamma_0(2)$ has the
special property of
preserving the parity of the denominator for rational $\nu=p/q$.  The
subgroup
$\Gamma(2)$, consisting of all elements of $\Gamma(1)$ which are congruent
to the
identity, mod 2, was also considered in \cite{Lutken} and has been further
investigated in
\cite{GW}.  Note however that there is no element of $\Gamma(2)$ which leaves
$\sigma^\ast
={{1+i}\over{2}}$ fixed, indeed there is no element of $\Gamma(2)$
which leaves
{\it any} $\sigma$ with $\infty>\Im(\sigma) > 0$ fixed.  
\par
It is a pleasure to thank Jan Pawlowski for discussions about the RG flow
of the quantum Hall effect.
\bigskip

\underbar{References}

\vfill\eject

\twocolumn[\hsize\textwidth\columnwidth\hsize\csname@twocolumnfalse%
\endcsname
\hbox{Table 1. Some examples of allowed transitions. The matrix $\gamma$
maps the points $\sigma=0$ and $\sigma=1$ to the transition}
\hbox{indicated in the leftmost column. Some representative experimental
support (not exhaustive) is also indicated.}
\hbox{The last two columns assume the semi-circle law ($\rho$ is the
resistivity).}
\smallskip
\hskip 1pt\vbox{\offinterlineskip
\hrule
\halign{&\vrule#&
  \strut\quad#\hfil\quad\cr
height 2pt&\omit&&\omit&&\omit&&\omit&&\omit&&\omit&\cr
&\vbox{\vskip .2cm\hbox{Transition}\vskip .2cm\hbox{$\nu_1\rightarrow\nu_2$}}
&&
\vbox{\hbox{\hskip .5cm$\gamma$}\vskip .2cm}&&
\vbox{\vskip .2cm\hbox{Critical}\vskip .1cm\hbox{Conductivity}}&&
\vbox{\vskip .2cm\hbox{Critical}\vskip .1cm\hbox{Resistivity}}&&
\vbox{\hbox{\hbox{$\sigma$} at \hbox{$\sigma^{Max}_{xx}$}}\vskip .1cm}&&
\vbox{\hbox{\hbox{$\rho$} at \hbox{$\rho^{Max}_{xx}$}}\vskip.1cm}&\cr
height 2pt&\omit&&\omit&&\omit&&\omit&&\omit&&\omit&\cr
\noalign{\hrule}
height 2pt&\omit&&\omit&&\omit&&\omit&&\omit&&\omit&\cr
&\hbox{${n+1 \rightarrow n}$}&&\hbox{$\left(\matrix{1&n\cr0&1\cr}\right)$}&&
\hbox{${(2n+1)+i\over 2}$}&&\hbox{${(2n+1)+i\over 2n^2+2n+1}^{(a)}$}&&
\hbox{${(2n+1)+i\over 2}$}&&\hbox{${(2n+1)+i\over 2n(n+1)}^{(b)}$}&\cr
&&&&&&&&&&&&\cr
&\hbox{${1\over 2n+1}\rightarrow 0$}&&\hbox{$\left(\matrix{1&0\cr2n&1\cr}
\right)$}&&
\hbox{${(2n+1)+i\over 2(2n^2+2n+1)}^{(c)}$}&&\hbox{$(2n+1)+i^{(d)}$}&&
\hbox{${1+i\over 2(2n+1)}$}&&\hbox{$(2n+1)+i\infty$}&\cr
&&&&&&&&&&&&\cr
&\hbox{${n\over 2n+1}\rightarrow{n+1\over 2n+3}$}&&
\hbox{$\left(\matrix{1&n\cr2&2n+1\cr}\right)$}&&
\hbox{${(4n^2+6n+3)+i\over 2(4n^2+8n+5)}$}&&
\hbox{${(4n^2+6n+3)+i\over 2n^2+2n+1}$}&&
\hbox{${(4n^2+6n+1)+i\over 2(2n+1)(2n+3)}$}&&
\hbox{${(4n^2+6n+1)+i\over 2n(n+1)}^{(e)}$}&\cr
&&&&&&&&&&&&\cr
&\hbox{${3n+2\over 4n+3}\rightarrow{3n+5\over 4n+7}$}&&
\hbox{$\left(\matrix{3&3n+2\cr4&4n+3\cr}\right)$}&&
\hbox{${(24n^2+58n+41)+i\over 2(16n^2+40n+29)}$}&&
\hbox{${(24n^2+58n+41)+i\over 18n^2+42n+29}$}&&
\hbox{${(24n^2+58n+29)+i\over 2(4n+3)(4n+7)}$}&&
\hbox{${(24n^2+58n+29)+i\over 2(3n+2)(3n+5)}^{(f)}$}&\cr
height 2pt&\omit&&\omit&&\omit&&\omit&&\omit&&\omit&\cr
}
\hrule}
{\hbox{(a) These points all lie on the semi-circle 
$\rho=i-\hbox{e}^{i\theta}$$
0\le\theta\le\pi$. For $n=1$ see \cite{TS4}.}}\hfill\break
{\hbox{(b) Assumes $n\ne 0$.}}\hfill\break
{\hbox{(c) These points all lie on the semi-circle 
$\sigma={1\over 2}(i-\hbox{e}^{i\theta})$, $0\le\theta\le\pi$.}}\hfill\break
{\hbox{(d) For $n=0$ see \cite{STSBC} and \cite{HS2TXM}, for $n=1$ see 
\cite{STSBC} and \cite{WJTP}, for $n=2$ see \cite{JWSTPW}.}}\hfill\break
{\hbox{(e) Assumes $n\ne 0$. For $n=1,\ldots,5$ and $n=-3,\ldots,-7$ see
\cite{DR}.}}\hfill\break
{\hbox{(f) For $n=0$ see \cite{DR}.}}\hfill\break]

\centerline{}
\vskip 2.4 cm
\epsfxsize=85mm
{\epsffile{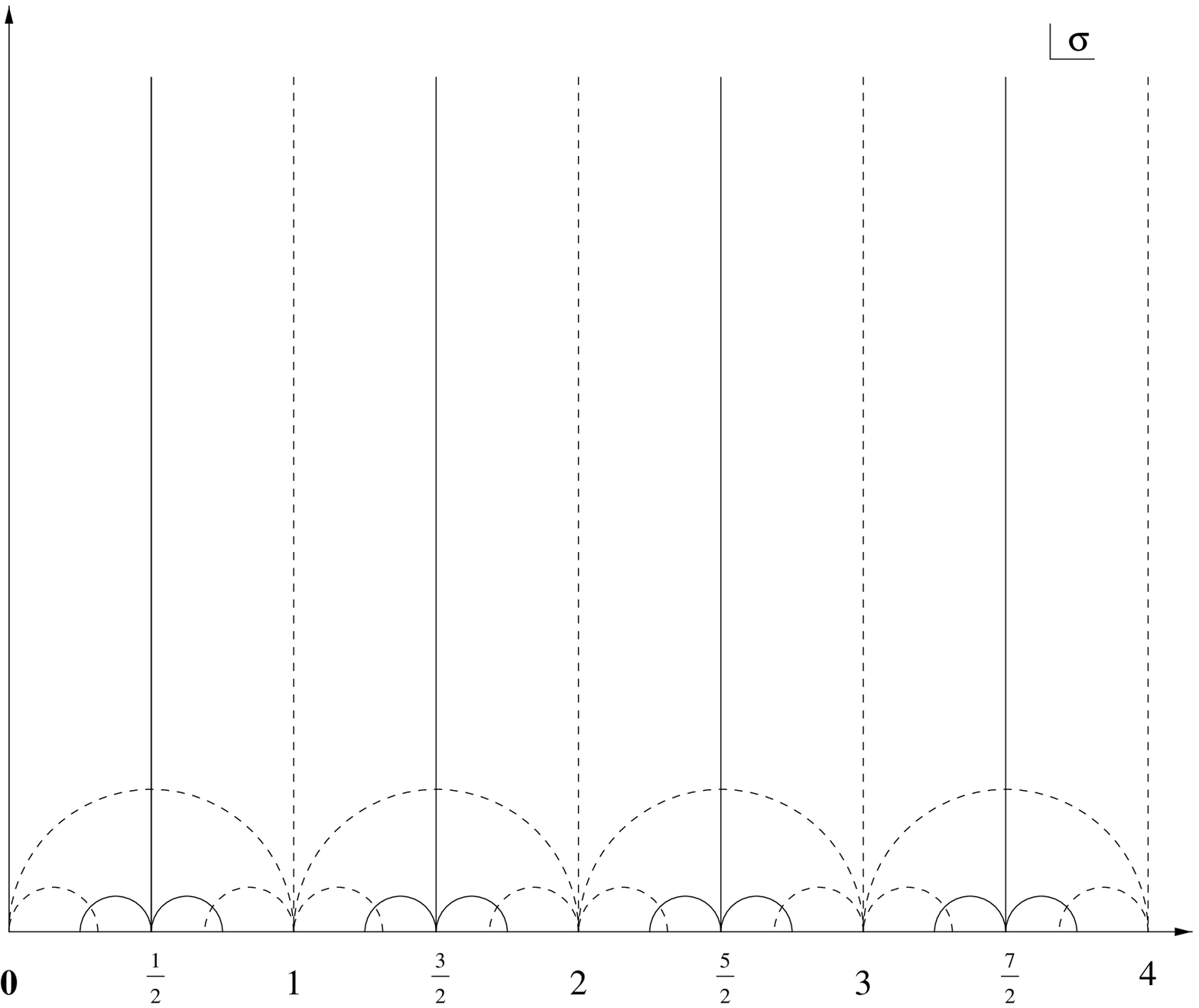}}
\vskip 1cm
Fig. 1 The phase structure in the upper-half complex
$\sigma$ plane. The solid curves represent phase boundaries and the dotted
curves allowed transitions. Points where dotted and solid lines cross are
critical points. 
\vskip 5cm
\epsfxsize=80mm
{\epsffile{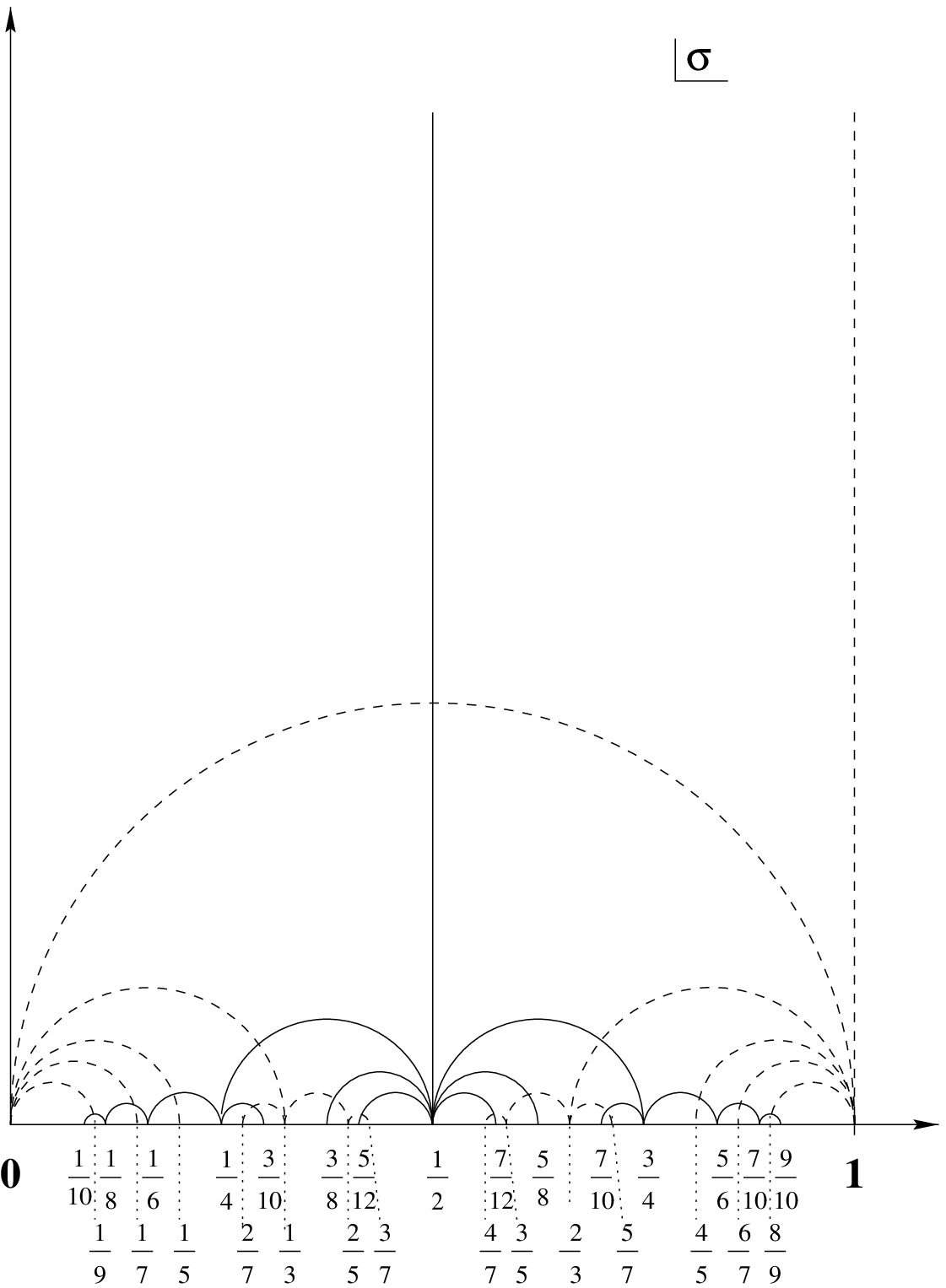}}
Fig. 2 A magnified view of the phase structure in the upper-half complex
$\sigma$ plane. 


\begin{references}
\bibitem{LutkenRossB} C. A. L\"utken and G. G. Ross, Phys. Rev. {\bf B48}, 2500
(1993)
\bibitem{Lutken} C. A. L\"utken, Nuc. Phys. {\bf B396}, 670 (1993);\hfill\break
J. Phys. A : Math. Gen. {\bf 26}, L811-L817 (1993)
\bibitem{BL}    C. P. Burgess and C. A. L\"utken, Nuc. Phys. {\bf B500},
367 (1997)\smallskip
\bibitem{KLZ}   S. Kivelson, D-H. Lee and S-C. Zhang, Phys. Rev. {\bf B46},
2223 (1992)\smallskip
\bibitem{LKZ}   D-H. Lee, S. Kivelson and S-C. Zhang, Phys. Rev. Lett.
{\bf 68}, 2386 (1992)\smallskip
\bibitem{Rankin} R. A. Rankin, {\sl Modular Forms and Functions}, C.U.P.
(1977)\smallskip
\bibitem{Khem}  D. E. Khmel'nitskii, Pis'ma Zh. Eksp. Teor. Fiz {\bf 38}, 454
(1983) 
        (JETP Lett. {\bf 38}, 552 (1983)\smallskip
\bibitem{WCTPR} H. P. Wei, A.M. Chang, D. C. Tsui, A. M. M. Pruisken and M.
Razeghi, Surf. Sci. {\bf 170}, 238 (1986)\smallskip
\bibitem{HHB}   Y. Huo, R. E. Hetzel and R. N. Bhatt, Phys. Rev. Lett.
{\bf 70}, 481 (1993)\smallskip
\bibitem{ZB}    J. Zang and J.L. Birman, Phys. Rev. {\bf B47}, 16305 (1993) 
\bibitem{WESTGE} R. Willett, J. P. Eisenstein, H. L. St\"ormer, D. C. Tsui,
A. C. Gossard and J. H. English, Phys. Rev. Lett. {\bf 59}, 1776 (1987)
\smallskip
\bibitem{LH}    R. B. Laughlin, Phys. Rev.
Lett. {\bf 50}, 1395 (1983)\hfill\break
        F. D. M. Haldane, Phys. Rev. Lett. {\bf 51}, 605 (1983)
\bibitem{CMH2F} R. G. Clark, J. R. Mallett, S. R. Haynes, J. J. Harris and
C. T.
Foxon, Phys. Rev. Lett. {\bf 60}, 1747 (1988)
\bibitem{DR}    A. M. Dykhne and I. M. Ruzin, Phys. Rev. {\bf B50}, 2369
(1994)\smallskip
\bibitem{RF} I. Ruzin and S. Feng, Phys. Rev. Lett. {\bf 74}, 154 (1995)
\smallskip
\bibitem{RCH} I.M. Ruzin. N.R. Cooper and B.I. Halperin, Phys. Rev.
{\bf B53}, 1558 (1996)\smallskip
\bibitem{CR}    J. L. Cardy and E. Rabinovici, Nuc. Phys. {\bf B205}, 1
(1982)
        J. L. Cardy, Nuc. Phys. {\bf B205}, 17 (1982)\smallskip
\bibitem{SW}    A. Shapere and F. Wilczek, Nuc. Phys {\bf B320}, 669 (1989)
\bibitem{LR}    C. A. L\"utken and G. G. Ross, Phys. Rev. {\bf B45}, 11837
(1992)
\bibitem{GW}    Y. Georgelin and J-C. Wallet, Phys. Lett. {\bf A224}, 303
(1997); 
Y. Georgelin, T. Masson and J-C. Wallet, J. Phys. {\bf A30}, 5065 (1997)
\bibitem{TS4}   D.~Shahar, D.~C.~Tsui, M.~Shayegan, E.~Shimshoni and
S.~L.~Sondhi,
 (cond-mat/9611011)
\bibitem{STSBC} D.~Shahar, D.~C.~Tsui, M.~Shayegan, R.~N.~Bhatt and
J.~E.~Cunningham,
Phys. Rev. Lett. {\bf 74}, 4511 (1995)\hfill\break
        D. Shahar, D. C. Tsui, M. Shayegan, J. E. Cunningham,
E. Shimshoni and
S. L. Sondhi, (cond-mat/9607127) Solid State Comm. {\bf 102}, 817 (1997)
\bibitem{HS2TXM} M. Hilke, D. Shahar, S. H. Song, D. C. Tsui, Y. H. Xie and
D. Monroe,
(cond-mat/9708239)
\bibitem{WJTP}  L. W. Wong, H. W. Jiang, N. Trivedi and E. Palm, Phys. Rev.
{\bf B51}, 18033 (1995)
\bibitem{JWSTPW} H. W. Jiang, R. L. Willett, H. L. Stormer, D. C. Tsui,
L. N. Pfeiffer
and K. W. West, Phys. Rev. Lett. {\bf 65}, 633 (1990)
\end{references}
\end{document}